# Three novel efficient Deep Learning-based approaches for compensating atmospheric turbulence in FSO communication system

M. A. Amirabadi,

Email: m_amirabadi@elec.iust.ac.ir

**Abstract-** One of the main problems encountered with Free Space Optical (FSO) Communication system is the atmospheric turbulence. Although many solutions exist for combating this effect, they have either high complexity or low performance. In this paper, a comprehensive investigation is developed, and three new effective Deep Learning (DL) based solutions are proposed. This paper, for the first time, deploys Deep Learning, transceiver learning, as well as transmitter learning in FSO communication. In addition, this is the first time that DL approach is implemented in FSO-Multi-Input Multi-Output (MIMO) communication. Results of the proposed structure are compared with the state of the art MQAM based FSO system with Maximum Likelihood Detection. Wide range of atmospheric turbulence, from weak to the strong regime, are considered; results indicate that the proposed structures despite less complexity, have the same performance as the outstanding conventional structure. In addition, in different MIMO structures (different combiners and the different number of transceiver apertures), the proposed structure still achieve the performance of the state of the art conventional system.

**Keywords-** Deep Neural Network, Constellation shaping, Free Space Optical communication, Multi-Input Multi-Output, Gamma-Gamma;

## I. Introduction

Free Space Optical (FSO) Communication system is the main competitor against conventional Radio Frequency Communication system, and it is widely applied in space and ground communications [1]. Because it has a huge un-licensed bandwidth, high data rate, high security, low cost, and low power consumption, as well as good performance and easy installation. Furthermore, FSO could be used at the bottleneck, and last-mile applications, and is a good candidate for the back-haul network of the next generation communication network [2].

Non-homogeneous disturbance of temperature and pressure over the atmosphere results in random changes in refractive index. Actually, the propagating media (atmosphere) induces phase disturbances and leads to intensity fluctuation, beam wandering and beam broadening of FSO signal. Among these effects, random fluctuations in the propagating signal intensity, which is called scintillation and is the most important effect of atmospheric turbulence. Various statistical distributions have been used in order to investigate the effects of atmospheric turbulences e.g., Exponential-Weibull [3], Generalized Malaga [4], Lognormal [5], Gamma-Gamma [6], and Negative Exponential [2]. Among them, Gamma-Gamma is mostly used, because it is highly accompanied with the actual results for weak to strong atmospheric turbulence intensities [7].

The Multi-Input Multi-Output (MIMO) technology characteristics have encouraged the researchers to incorporate multiple transmit and receive apertures in FSO communication [8]. Actually, one of the promising techniques to combat atmospheric turbulence effects is the FSO-MIMO structure. Compared with FSO Single-Input Single-Output (SISO), this technology has improved capacity, diversity gain, and coverage area. It should be noted that diversity is a widely used efficient technique for improving the performance of FSO communication system [9]. In this technique, different copies of the original signal are transmitted, encountered with different fading channels, and collected at the receivers. Combination of the received copies could really help to better recovery the transmitted signal [10]. Various combining schemes are available, which among them Maximum Ratio Combiner (MRC), Equal Gain Combiner (EGC), and Selection Combiner (SC) are some of the well-known combiners [11].

Considering the perfect prediction of the statistical model of atmospheric turbulence, the Maximum Likelihood (ML) detector has been widely used in the FSO system. ML is the optimal detection method but has high complexity of calculating $n$-dimensional integrals for each of $2^n$ bit sequences, which is the main barrier of its practical implementation. Although sub-optimal and low-complexity extensions of ML have been developed, they show reduced performance. Hence, it's meaningful to investigate detection techniques with low-complexity that could maintain performance without any implementation barrier [12].

Machine Learning is the powerful interconnection between mathematics, computer, and biology, which recently has been successfully applied in different branches of science, especially Optical Communication (OC) including Visible Light Communication (VLC), Fiber OC, and FSO communication, in both network and physical layer perspectives. As a core member in the Machine Learning community, DL (in special, Deep Neural Network (DNN)) showed significant performance in OC applications. The DL offers powerful statistical signal processing tools that can be applied for accurate amplitude and phase noise characterization. The significant advantage is that physics of the channel, network, or device can be included in the characterization and demodulation of the DL algorithms. Furthermore, it can be used for non-white and also non-Gaussian noise. DL could learn the received data impairments and generate an accurate probabilistic model for impairments [13].

Among DL algorithms, DNN is the most widely used technique in OC, which has proved to be sufficient alternatives to conventional numerical, analytical, or empirical methods. DNN is simple and have low complexity, and can model complex multi-dimensional nonlinear relationships; it is generic and its response is fast. Due to these advantages, DNN attracted considerable attention and became a powerful tool in OC perspectives such as Fiber OC, FSO, or VLC. DNN first learns the model of the signal/system behavior and can be used in high-level simulation and design, providing fast answers. DNN learns the input-output data relationship by using several hidden layers, which each consisted of multiple connected neurons by some weights, biases and activation functions that represent the importance of each connection [14].

Machine Learning has been widely applied to the OC systems for various purposes, such as fiber effects mitigation [15], performance monitoring [16], modulation format identification [17] and OC network applications [18]. However, most of the investigations are developed based on Fiber OC; few works considered wireless OC applications. It should be noted that even the few existing works on FSO investigation are very basic. For example, the |Support Vector Machine is used in [12] as a detector to combat the noise and scintillation effects, or a backpropagation Artificial Neural Network is applied in [1] for correcting distortion in a sensor-less adaptive optic system. In [19], Conventional Neural Network is used as a demodulator for a turbo-coded orbital angular momentum-shift keying FSO system at strong atmospheric turbulence, or in [20] (as an extension of [19]), CNN is used for joint atmospheric turbulence detection and demodulation for an orbital angular momentum FSO system. Compared to previous approaches using the self-organizing mapping, and other CNNs, this method achieved a higher atmospheric turbulence detection accuracy. In [21] some decision tree based Machine Learning methods including Ada Boost Classifier with Decision Tree Classifier, Random Forest Classifier, and Gradient Boosting Classifier are used in order to estimate received optical power parameter based on series of weather parameters in a hybrid FSO/RF system with hard switching.

Previously, Machine Learning was used as a tool only at the receiver side, and this point was ignored that Machine Learning is a tool and could be used anywhere. However, recently, some investigations in Fiber OC implemented End-to-End implementation of DNN in the OC system, and jointly optimized the transceiver parameters. This technique is used for auto encoding (and combating dispersion) [22-23] or constellation shaping (and combating nonlinearity) [24, 25] in Fiber OC. However, one of the main drawbacks of training an End-to-End DNN in Fiber OC is its requirement to a channel model for training the transmitter. But there is no accurate model for Fiber OC, and a roughly model should be used; actually, this causes that the transmitter could not perform at its best in real conditions. However, this drawback does not exist in FSO system, because FSO channel models (presented in literature) is highly accompanied with the actual results; so, End-to-End DNN works in FSO system as it should work.

In this paper, a comprehensive investigation is developed, and three DNN based transceiver structures are presented for combating the effect of atmospheric turbulence. In the first structure, DNN is implemented at the transmitter and works as a constellation shaper, in the second structure, DNN is implemented at the receiver and detects the received symbol. In the third structure, an Ent-to-End DNN is deployed to jointly shape constellation, and detect the received symbol. In order to have better results, the MIMO implementation of the proposed structures is also considered. In order to have a comprehensive investigation, performance of the proposed structures is investigated in both SISO, and MIMO structure for different modulation orders, different atmospheric turbulence regimes (Gamma-Gamma model, from weak to strong), and different combining schemes (MRC, EGC, and SC). According to the statement of the three last paragraphs, and to the best of the authors' knowledge, novelties and contributions of this work that are done for the first time in Machine Learning for FSO investigations include:

1- Deploying DNN in the FSO system.
2- Considering MIMO structure in Machine Learning for FSO investigations (and consequently considering MRC, EGC, and SC in Machine Learning for FSO investigations).

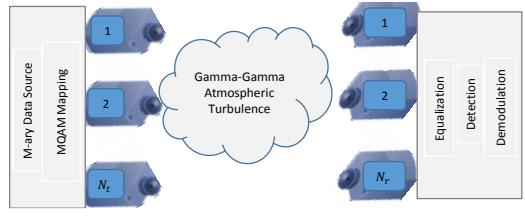

Fig.1. MQAM based FSO-MIMO communication system with ML detection.

3- Deploying transmitter learning, as well as transceiver learning in the FSO system.
4- Considering a wide range of atmospheric turbulence from weak to strong in Machine Learning for FSO investigations.
5- Presenting Machine Learning based joint constellating shaping, and detecting technique for FSO system.
6- Developing an accurate End-to-End DNN in OC (because in spite of Fiber OC, FSO channel models are accurate and have high accompany with the actual results).

The rest of this study is organized as follows; in order to reduce the complexity, and making paper better understanding and tracking, discussions about FSO communication and Machine Learning are separated to sections II, and II for FSO channel and system model, respectively, and section IV for Machine Learning. In addition, sections II, and V (simulation results and discussions section) are divided into SISO, and MIMO subsections. However, since the used Machine Learning methods and channel model are the same in SISO, and MIMO structures, sections II, and IV are not sub-divided. Section VI is the conclusion of this work.

## II. FSO Channel Model

Even in perfect FSO channel condition, (i.e., clear weather with 0.5 dB/km attenuation), still, the major challenge faced by the FSO system is the atmospheric turbulence. The most important impact of atmospheric turbulence is the intensity fluctuation of the received signal, which is called scintillation effect. The most important parameter characterizing the strength of atmospheric turbulence is refractive index structure parameter ($c_n^2$), which is a function of geographical location, altitude and time.

The temperature and pressure gradients near the earth surface are higher from other locations. Accordingly, $c_n^2$ at the sea level has the highest value and decreases while rising the altitude. One of the most popular formulations which define $c_n^2$ parameter is Hufnagel Valley model:

$$c_n^2 = 0.00594 \left(\frac{v}{27}\right)^2 (10^{-5}h)^{10} \exp\left(-\frac{h}{1000}\right) + 2.7 \cdot 10^{-16} \exp\left(-\frac{h}{1500}\right) + A_0 \exp\left(-\frac{h}{100}\right), \quad (1)$$

where parameter $h\ [m]$ refers to altitude, $v\ [m/s]$ is wind speed, and $A_0$ is a turbulence strength parameter, which at earth surface is $\sim 1.7 \cdot 10^{-14}\ m/s^{-2/3}$. Usually, (close to earth surface,) $c_n^2$ is in the range from $10^{-15}\ m^{-2/3}$ for weak turbulences to $10^{-13}\ m^{-2/3}$ for strong turbulence. It turns out that FSO link applications employ long horizontal atmospheric path with relatively long distances. Thinking about these kinds of applications $c_n^2$ parameter is considered quasi-constant. The Scintillation index could be calculated from the following formula [21]:

$$\sigma_R^2 = 1.23 c_n^2 k^{7/6} l^{11/6}, \quad (2)$$

where $k = 2\pi/\lambda$ is wave number, $l[m]$ is FSO link distance. There are various models available for atmospheric turbulence, however, among them, Gamma-Gamma is mostly used and is highly accompanied with actual results for weak to strong regimes. Assuming plane wave propagation for FSO signal, the shaping and scaling parameters ($\alpha$ and $\beta$) that characterize the irradiance fluctuation in Gamma-Gamma model are related to the atmospheric conditions by [6]:

$$\alpha = \left[\exp\left(0.49\sigma_R^2/\left(1 + 1.11\sigma_R^{12/5}\right)^{7/6}\right) - 1\right]^{-1}, \quad (3)$$

$$\beta = \left[\exp\left(0.51\sigma_R^2/\left(1 + 0.69\sigma_R^{12/5}\right)^{5/6}\right) - 1\right]^{-1}. \quad (4)$$

Considering these parameters, the probability distribution function of Gamma-Gamma atmospheric turbulence becomes as follows:

$$f(I) = \frac{2(\alpha\beta)^{\frac{\alpha+\beta}{2}}}{\Gamma(\alpha)\Gamma(\beta)} I^{\frac{\alpha+\beta}{2}-1} K_{\alpha-\beta}(2\sqrt{\alpha\beta I}); \quad I > 0 \qquad (5)$$

where $I$ is the atmospheric turbulence intensity, $\Gamma(.)$ Is the well-known gamma function, $K(.)$ is Modified Bessel function of the second kind.

### III. FSO System Model

In this paper, four main structures are proposed, which three of them are the novelties of this paper. In addition, another novelty of this paper is presenting MIMO structure in Machine Learning for FSO communication. So, in order to make tracking and understanding of this paper easier, this part is divided into SISO, and MIMO sub-sections. However, the only difference between the proposed four structures is related to the use of DNN in their transceivers. So, in optics communication perspective, they are the same. Therefore, this section deploys mathematics based on simple SISO, and MIMO FSO structures (see Fig.1. for MIMO).

#### A. SISO

Consider $x$ as the source symbol mapped on MQAM (or DNN constructed) constellation, these symbols are first carried on an electrical signal, and then the electrical signal is converted to an optical by an optical modulator and finally transmitted through the FSO channel. At the channel, the FSO signal encounters with the multiplicative atmospheric turbulence, and at the receiver is added by additive white Gaussian noise (AWGN) with zero mean and $\sigma^2$ variance. At the receiver, the optical signal is converted back to an electrical signal by a photodetector with conversion efficiency if $\eta$. The electrical signal is then sampled and the symbol at the ML (or DNN based) detector becomes as follows:

$$y = \eta(Ix + n), \qquad (6)$$

where $n$ is AWGN at the receiver input.

#### B. MIMO

Consider $x$ as the source symbol mapped on MQAM (or DNN constructed) constellation; $N_t$ copies of the modulated optical signal are transmitted through parallel $N_t$ FSO transmitters, encountered with FSO-MIMO channel. Then FSO signals are received by $N_r$ photodiodes, added by AWGN, and converted to an electrical signal by conversion efficiency of $\eta$. So, there are $N_r$ signals which should be combined, sampled, and fed into maximum-likelihood likelihood (or DNN based) detector. The received symbol of the $i - th, i = 1, ..., N_r$ receive aperture is as follows [26]:

$$y_i = \sum_{j}^{N_t} I_{i,j} x + n_i, \qquad (7)$$

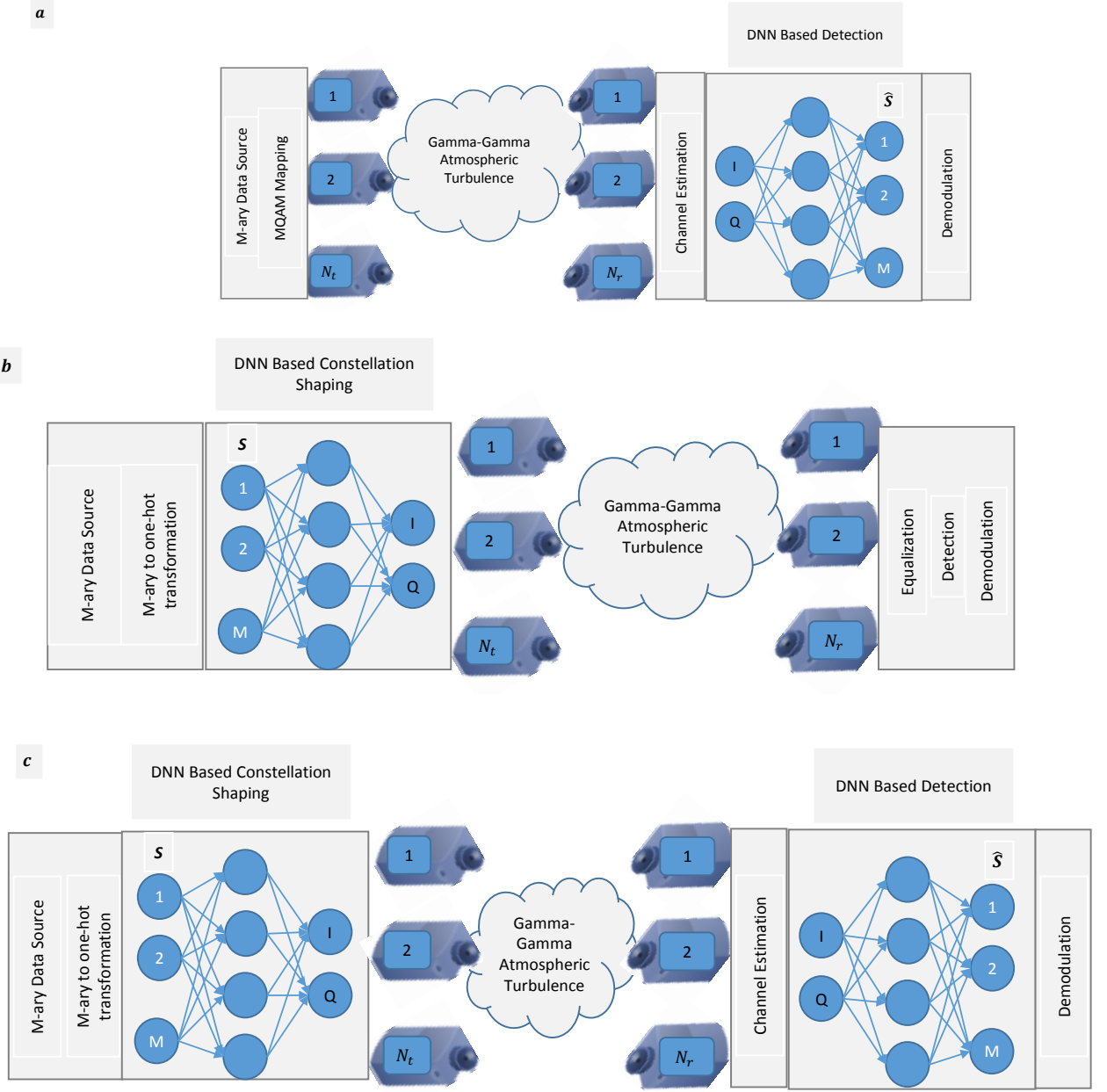

Fig.2. a, b, c. Three proposed DNN based FSO-MIMO structures.

where $n_i$ is AWGN with zero mean and variance $\sigma^2$; $I_{i,j}$ is the atmospheric turbulence intensity of the link between the $j-th$ transmit aperture and $i-th$ receive aperture.

In this paper, in order to have a comprehensive investigation over FSO-MIMO structure, SC, EGC, and MRC are implemented. The SC scheme selects the received signal with maximum SNR; assuming $p$ as the index of the selected receive aperture, the maximal-likelihood detector for the SC will be as follows:

$$\hat{x} = \min_{\tilde{x}} \left| y - \eta \sum_{j=1}^{N_t} I_{p,j} \tilde{x} \right|^2, \tag{8}$$

where $\tilde{x}$ is a symbol of MQAM (DNN constructed) constellation. EGC multiplies the received signals by the phase conjugate of the channel intensity (to remove channel phase), and then sums them. However, FSO channel intensity is real, therefore, EGC output becomes as follows [26]:

$$y = \sum_{i=1}^{M} y_i = \eta \sum_{i=1}^{M} \sum_{j}^{N_t} I_{i,j} x + \sum_{i=1}^{M} n_i. \tag{9}$$

ML Receiver for the EGC is given by [26]:

$$\hat{x} = \min_{\tilde{x}} \left| y - \eta \sum_{i=1}^{N_r} \sum_{j=1}^{N_t} I_{i,j} \tilde{x} \right|^2. \tag{10}$$

The combined electrical signal in MRC receiver is as follows [26]:

$$y = \sum_{i=1}^{N_r} \sum_{j=1}^{N_t} I_{i,j} y_i = \eta^2 \sum_{i=1}^{N_r} \left( \sum_{j=1}^{N_t} I_{i,j} \right)^2 x + \eta \sum_{i=1}^{N_r} \sum_{j=1}^{N_t} I_{i,j} n_i. \tag{11}$$

ML Receiver for the MRC is given by [26]:

$$\hat{x} = \min_{\tilde{x}} \left| y - \eta^2 \sum_{i=1}^{N_r} \left( \sum_{j=1}^{N_t} I_{i,j} \right)^2 \tilde{x} \right|^2. \tag{12}$$

## IV. Proposed DNN based structures

In this section, the main focus is on the Machine Learning perspective of the proposed structures. According that the proposed structures use the same DNN structure, and in order to avoid confusion, only one of the three DNN based structures is explained (Fig.2.c, because it contains the structure of the two others); the required explanation for the two other structures could be extended realized considering sections II, and III. The proposed structures include 1-QAM constellation-DNN based detection (QAM-DNN), 2- DNN based constellation shaper-ML detection (DNN-ML), 3- DNN based constellation shaper-DNN based detection (End-to-End DNN) (see Fig. 2. a, b, c, respectively). This section does not focus on the communication theory; accordingly, both SISO and MIMO structures have the same procedure and are not separated in this section.

The processing of implemented conventional transceivers (transmitter of the Fig. 2.a, and receiver of Fig. 2.b) are explained in section II. In this section, the main focus is on processing in DNNs, therefore, only the third structure (Fig.2.c) is discussed in this section. Because the other structures are a combination of DNN and conventional system, and their processing can be understood while considering explanations of sections II, and III. So, in this section focus is on Fig.2.c.

The generated $N_{sym}$ digital M-ary symbols are first converted to $N_{sym}$ one-hot vectors, each of size $M$. The produced one-hot vectors are entered a DNN with $M$ input and 2 output neurons. The number of hidden layers is $N_{hid}$, the number of neurons per hidden layer is $N_{neu}$, the activation functions are $\boldsymbol{\alpha}(.)$, weights and biases are $\boldsymbol{W}$ and $\boldsymbol{b}$, respectively. Complex summation of the output of the DNN results in a complex number which stands for the location of the transmitted symbol in the signal constellation. The purpose is to adjust DNN parameters such that the effect of atmospheric turbulence on the propagating signal is minimized. The DNN output is either copied (for MIMO case) or not (for SISO case) and transmitted through FSO SISO/MIMO channel. At the receiver, AWGN with zero mean and $\sigma^2$ variance is added to the signal. The received signal is converted to an electrical signal by conversion efficiency of $\eta$, and in MIMO case combined by MRC/EGC/SC shame, and fed to a DNN.

So, there is a complex signal at the DNN output. The real and imaginary parts of this signal are separated and fed to the DNN inputs. The receiver side DNN has exactly the inverse structure of the transmitter side DNN. The task of the DNN at the receiver is to recover original digital M-ary symbols.

In order to solve this problem efficiently, end to end training is implemented; i.e. the parameters of transceiver side DNNs are adjusted simultaneously. The first step in training a DNN is to select and tune its hyperparameters, which include sample size to batch size ratio, layer type, number of layers, number of neurons, activation function, cost function, optimizer, learning rate, and number of iterations. Sample size to batch size ratio is important because entering the whole data at once into the DNN leads to underfitting while dividing it into several batches helps DNN to better understand the data structure. There are several layer types in DNN, which each of them is proper for one or some tasks, and are dependent on the input data type, e.g. Recurrent Neural Network is proper when data has passed a memory channel, or Convolutional Neural Network is preferable for imagery data sets. The number of layers, as well as neurons, should be adjusted by trial and test, and there is no specific rule for tuning them. For selecting the activation function, there is a long story, activation functions extended during the time and according to complexity, accuracy, and timing demands, there is a tradeoff between them, however some of them such as tanh, sigmoid, relu are shown to be proper for Machine Learning for OC applications, and frequently used in literatures. After defining these hyperparameters, its turn to justify the cost function.

After selecting and tuning the hyperparameters, it is turn to train the DNN. The inputs of each layer are multiplied by weights, added by biases, summed, and then entered activation function (neuron). This procedure at all defined layers of the proposed End-to-End DNN based on Fig.2.c. The aim is to reduce the difference between input and output one-hot vectors ($s$, and $\hat{s}$). Therefore a loss function should be defined and calculated for each individual transmitted symbol and expected over whole batch size. The proposed loss function could be defined as follows [24]:

$$L(\boldsymbol{\theta}) = \frac{1}{K}\sum_{k=1}^{K}\left[l^{(k)}(\boldsymbol{s}, \hat{\boldsymbol{s}})\right] \tag{13}$$

where $\boldsymbol{\theta}$ is the DNN parameter vector (include weights and biases), $K$ is the batch size, $l(.,.)$ is loss function. Then it's the turn of the most important step in DNN; training aims to adjust DNN parameters to minimize the loss function. This could be done iteratively by the following formulation:

$$\boldsymbol{\theta}^{(j+1)} = \boldsymbol{\theta}^{(j)} - \eta \nabla_{\boldsymbol{\theta}} \tilde{L}(\boldsymbol{\theta}^{(j)}) \tag{14}$$

where $\eta$ is learning rate (which if set too large causes an unstable system and if set too low means slow convergence), $j$ is the training step iteration, and $\nabla_{\boldsymbol{\theta}} \tilde{L}(.)$ is the estimate of the gradient. Actually, the error derivation ($\nabla_{\boldsymbol{\theta}} \tilde{L}(\boldsymbol{\theta}^{(j)})$) is fed back to the DNN as an updating guide; the positive step size is known as the learning rate ($\eta$) [14]. Optimization is a tricky subject, which depends on the input data quality and quantity, model size, and the contents of the weight matrices. A lot of optimizers are tuned for rather specific problems like image recognition or ad click-through prediction. Generally, trial and error is the way to determine the best optimizer. Stochastic Gradient Descent methods could be used for determining update direction and solving the above problem. Among them, Adam is the most widely used algorithm for this task [22].

### V. Simulation results and discussions

In this section, simulation results for SISO, and MIMO structures are gathered separately. Simulations for conventional structures are done in MATLAB, and simulations for DNN based structures are done in Tensorflow environment. Gamma-Gamma and Gaussian random variables are both available in MATLAB and Tensorflow. Table. 1. Shows the used hyperparameters in simulations. The selection of hyperparameters is done roughly by manually selecting some parameters and observing the results. Considering sample size to batch size ratio equal to 4 has better performance in tuning. The layer types are deep artificial neural networks, which are selected due to the application type, this type can approximate any arbitrary, nonlinear, continuous, multidimensional function. There are many activation functions available in Tensorflow, which are listed in Table.2, e.g., sigmoid, and hyperbolic tangent, which are bounded, continuous, monotonic and continuously differentiable. However, these functions contemplate all neurons and have high complexity. Elu community such as Relu, Relu6, Crelu, and Elu have lower complexity. In hyperparameter tuning, Crelu for MIMO, and Relu for SISO showed better performances [23]. The number of hidden neurons depends highly on the degree of nonlinearity and model dimensionality of the model. Highly nonlinear systems require more neurons, while smoother systems require fewer neurons. Tuning indicates that $N_{hid} = 40$ neuron in each hidden layer has better performance [14]. After designing the DNN, it is the turn of selecting a loss function, MMSE, and Cross entropy are the well-known loss functions, which Cross entropy is mostly used in DNN for OC applications, and therefore is used in this paper. The last step it choosing the optimizer; among the optimizers available in Tensorflow (see Table.3.) Adam is the mostly used one and shows better performance in the tuning procedure. Tuning indicates that the iteration number of 1000 could bring the required accuracy.

Table.1. Tuned hyperparameters in this paper.

| Hyperparameter | Value |
|---|---|
| Modulation order | 16/4 |
| Number of layers | 4 |
| Number of hidden neurons | 40 |
| Batch size | 16*256*16 |
| Sample size/batch size | 4 |
| Number of Iterations | 1000 |
| Activation function | Relu/Crelu |
| Loss | Softmax cross entropy with logits_v2 |
| Optimizer | Adam |
| Learning rate | 0.005 |
| Gamma-Gamma Atmospheric turbulence Intensity | Strong ($\alpha = 4.2, \beta = 1.4$)/ Moderate ($\alpha = 4, \beta = 1.9$)/ Weak ($\alpha = 11.6, \beta = 10.1$) |

Table.2. Available activation functions in Tensorflow.

| Activation function | Equation |
|---|---|
| Relu | $f(x) = \begin{cases} x & x > 0 \\ 0 & x \leq 0 \end{cases}$ |
| Leaky Relu | $f(x) = \begin{cases} x & x > 0 \\ \alpha x & x \leq 0 \end{cases}$ |
| Crelu | $f(x) = \begin{cases} max(0, x) & x > 0 \\ max(0, -x) & x \leq 0 \end{cases}$ |
| Elu | $f(x) = \begin{cases} x & x > 0 \\ \alpha(e^x - 1) & x \leq 0 \end{cases}$ |
| Selu | $f(x) = \lambda \begin{cases} x & x > 0 \\ \alpha(e^x - 1) & x \leq 0 \end{cases}$ |
| Relu-6 | $f(x) = \begin{cases} 6 & 6 < x \\ x & 0 < x \leq 6 \\ 0 & x \leq 0 \end{cases}$ |
| Tanh | $f(x) = \dfrac{e^x - e^{-x}}{e^x + e^{-x}}$ |
| Sigmoid | $f(x) = \dfrac{1}{1 + e^{-x}}$ |
| Softmax | $f(x_i) = \dfrac{e^{-x_i}}{\sum_j e^{-x_i}}$ |
| Softsign | $f(x) = \dfrac{x}{1 + |x|}$ |
| Softplus | $f(x) = \ln(1 + e^x)$ |

Table.3. Available optimizers in Tensorflow.

| Optimizer |
|---|
| Adagrad Optimizer |
| Adadelta Optimizer |
| Adam Optimizer |
| Proximal Adagrad Optimizer |
| Gradient Descent Optimizer |
| Proximal Gradient Descent Optimizer |
| Ftrl Optimizer |
| Momentum Optimizer |
| RMS Prop Optimizer |

A. **SISO**

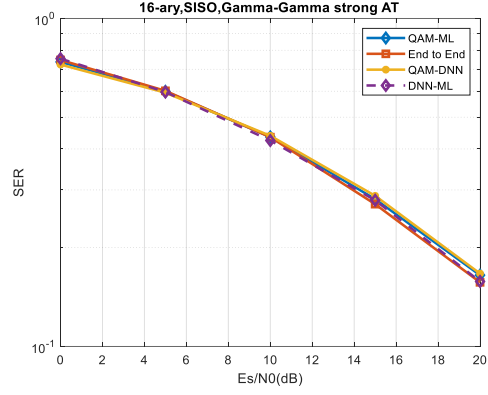

Fig.3. SER of FSO-SISO structure as a function of Es/N0, for the strong regime of atmospheric turbulence, for QAM-ML and End-to-End, DNN-ML, and QAM-DNN transceivers.

In Fig.3, SER of FSO-SISO structure is plotted as a function of Es/N0, for the strong regime of atmospheric turbulence ($\alpha = 4.2, \beta = 1.4$), for QAM-ML and End-to-End, DNN-ML, and QAM-DNN transceivers. It can be seen that end to end DNN performs the same as DNN-ML transceiver, as well, QAM-DNN matches QAM-ML, which indicate the efficiency of the proposed DNN based receivers. In addition, DNN-ML outperforms QAM-ML, which indicates the efficiency of the DNN based transmitter.

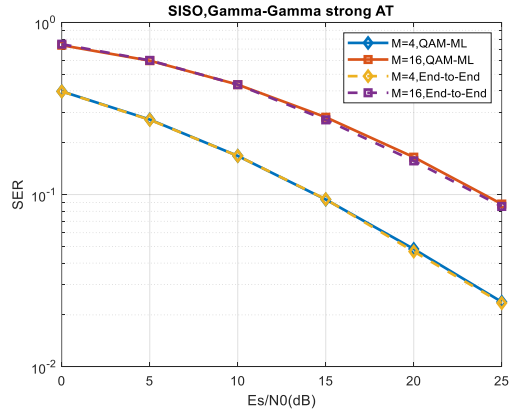

Fig.4. SER of FSO-SISO structure as a function of Es/N0, for different modulation orders, for the strong regime of atmospheric turbulence, for both QAM-ML and End-to-End DNN transceivers.

In Fig.4, SER of FSO-SISO structure is plotted as a function of Es/N0, for different modulation orders, for the strong regime of atmospheric turbulence, for both QAM-ML and End-to-End DNN transceivers. As can be seen, End-to-End DNN has the same performance as the state of the art QAM-ML based transceiver in $M = 4, 16$. The significance of obtained results could be understood while noting that End-to-End DNN has low complexity compared with QAM-ML based transceiver, and its practical implementation has no barrier. The interesting point in the obtained results is that by a simple hyperparameter tuning, and with the use of a simple structure, could achieve the performance of the state of the art MQAM-ML structure.

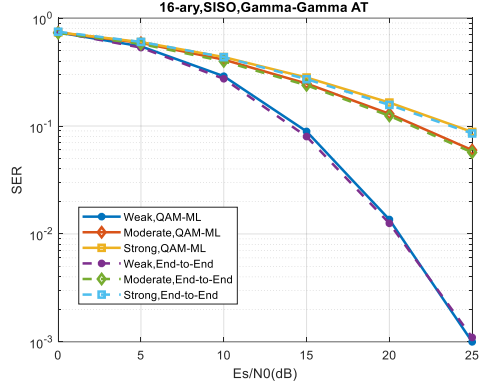

Fig.5. SER of the proposed FSO-SISO structure as a function of Es/N0, for different atmospheric turbulence regimes, for both 16-ary QAM-ML and End-to-End DNN transceivers.

In Fig.5, SER of the proposed FSO-SISO structure as a function of Es/N0, for different atmospheric turbulence regimes, for both 16-ary QAM-ML and End-to-End DNN transceivers. It can be seen that even in a complicated 16-ary constellation, the proposed End-to-End DNN structure achieve the performance of state of the art QAM-ML structure at all atmospheric turbulence regimes from weak to strong. In addition, it should be considered that implementing a better hyperparameter tuning will definitely improve the performance and increase the outperforming over state of the art QAM-ML based structure.

B. MIMO

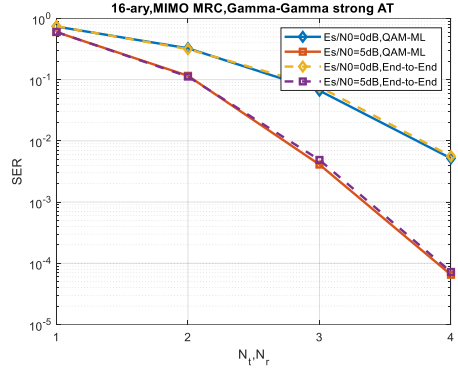

Fig.6. SER of FSO-MIMO structure as a function of Es/N0, for the different number of transceiver apertures, for the strong regime of atmospheric turbulence, for QAM-ML and End-to-End, transceivers.

In Fig.6, SER of FSO-MIMO structure is plotted as a function of Es/N0, for the different number of transceiver apertures, for the strong regime of atmospheric turbulence, for QAM-ML and End-to-End DNN transceivers. It can be seen that at different numbers of transceiver apertures, the proposed end to end structure has the same performance as the state of the art QAM-ML structure, and this shows that this structure is robust to any changes in the structure of the system, and by a simple training procedure could get adapted to the any new structure. So, End-to-End DNN, due to its lower complexity is a good choice for transceivers that have a variable number of active antennas. Because in traditional MIMO structures with a variable number of active antennas, several transceiver processing, each related to a specific design should be implemented and this increases the cost, and volume; however, the proposed structure easily solves this problem, and reduces complexity, cost, and volume, while maintaining performance and accuracy.

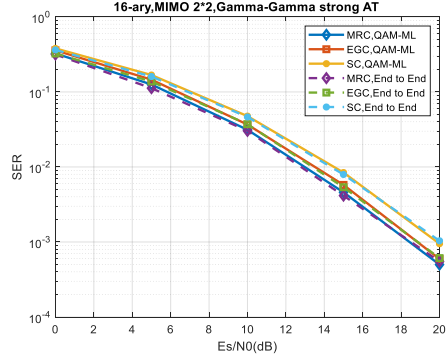

Fig.7. SER of the proposed FSO-MIMO structure as a function of Es/N0, for different combining schemes, for both 16-ary QAM-ML and End-to-End DNN transceivers, for strong atmospheric turbulence regime.

In Fig.7. SER of the proposed FSO-MIMO structure is plotted as a function of Es/N0, for different combining schemes, for both 16-ary QAM-ML and End-to-End DNN transceivers, for strong atmospheric turbulence regime. As can be seen, MRC in both QAM-ML and End-to-End DNN structures outperforms EGC, and EGC outperforms SC. In addition, the same as before, End-to-End DNN performs the same as QAM-ML structure for different combining schemes. However, this point should be considered that number of transceiver apertures is only 2, and the investigation is done over strong atmospheric turbulence regime; therefore, MRC, EGC, and SC are not so much different from each other. However, an increasing number of apertures would increase this difference.

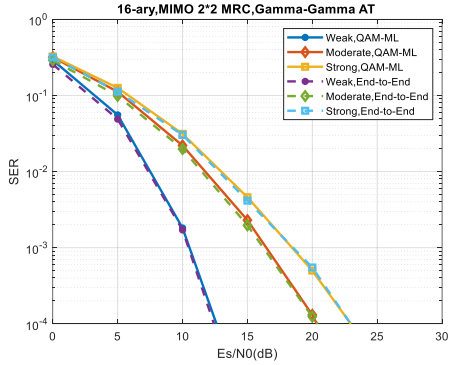

Fig.8. SER of the proposed FSO-MIMO structure as a function of Es/N0, for different atmospheric turbulence regimes, for both 16-ary QAM-ML and End-to-End transceivers, for MRC $2 \times 2$ structure.

In Fig.8. SER of the proposed FSO-MIMO structure is plotted as a function of Es/N0, for different atmospheric turbulence regimes, for both 16-ary QAM-ML and End-to-End transceivers, for MRC $2 \times 2$ structure. As can be seen, end to end DNN could achieve performance of the state of the art QAM-ML structure whiles in SISO system, it outperforms QAM-ML; there are several reasons for it, first that the hyperparameters of end to end DNN are first tuned for SISO structure, and second is that MIMO structure has more complexity, and learning its structure requires more complexity compared with SISO system. Although hyperparameter tuning for MIMO is done based on previous knowledge, it presents a significant performance similarity at all atmospheric turbulence regimes.

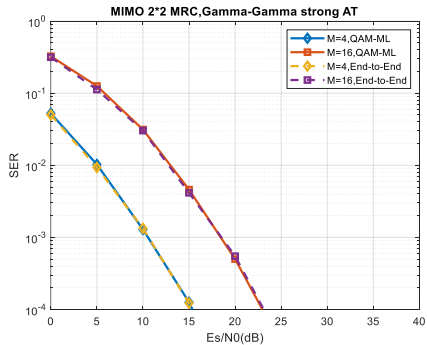

Fig.9. SER of the proposed FSO-MIMO structure as a function of Es/N0, for different modulation orders, for both QAM-ML and End-to-End transceivers, for MRC 2 × 2 structure, for strong regime of atmospheric turbulence.

In Fig.9, SER of the proposed FSO-MIMO structure as a function of Es/N0, for different modulation orders, for both QAM-ML and End-to-End transceivers, for MRC 2 × 2 structure, for strong regime of atmospheric turbulence. As can be seen, the same as before, end to end DNN can perfectly learn the data structure and presents the same performance the same as the state of the art QAM-ML based structure. Compared with SISO structure, changing modulation order in MIMO structure bring more gain and as can be seen, MIMO structures present favorable performance even at low Es/N0 regimes.

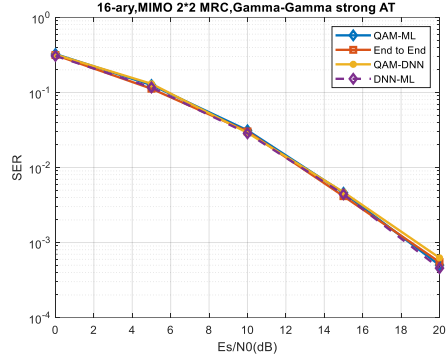

Fig.10. SER of FSO-MIMO structure as a function of Es/N0, for QAM-ML, End-to-End DNN, DNN-ML, and QAM-DNN transceivers, for strong regime of atmospheric turbulence, for MRC 2 × 2 structure.

In Fig.10, SER of FSO-MIMO structure is plotted as a function of Es/N0, for QAM-ML, End-to-End DNN, DNN-ML, and QAM-DNN transceivers, for strong regime of atmospheric turbulence, for MRC 2 × 2 structure. As can be seen, End-to-End DNN achieves the performance of DNN-ML, and the same thing is true for QAM-DNN and QAM-ML. However, QAM-ML and DNN-ML have a very low difference, because of implementation of DNN in one side rather than both sides, but because of long iteration number and using a long sample size for learning the proposed DNN based structures, they have almost the same performance. As a final note, it should be emphasized that although hyperparameter tuning improves performance, it's complicated and time and power consuming. In addition, the improvements achieved in hyperparameter tuning are not so much considerable, that deserve adding complexity and processing to achieve them. Because as can be seen in the results, by roughly tuning, proposed DNN based structures could achieve the performance of the state of the art conventional systems.

## VI. Conclusion

In this paper three novel Machine Learning based FSO-MIMO structures are proposed as solutions for combating effects of atmospheric turbulence on the performance of FSO system. In these structures, for the first time, transceiver learning, transmitter learning as well as DL are deployed in Machine Learning for FSO investigations. In addition, this paper for the first time considered a MIMO structure in Machine Learning for FSO investigations. In order to have a comprehensive investigation, a wide range of atmospheric turbulences, from weak to strong and different MIMO combining scenarios are considered. In order to show the efficiency of the proposed structures, results are compared with the well-known MQAM based FSO-MIMO system with ML detection. Results indicate that despite less complexity of the proposed structures, the proposed structure could perform as well as the mentioned state of the art structure. In addition, this structure is robust to any changes in the structure of the system, and channel, and by a simple training procedure could get adapted to any structural changes.